\documentclass[9pt,conference]{IEEEtran}
\IEEEoverridecommandlockouts
\usepackage{amsmath,amssymb,amsfonts}
\usepackage{algorithmic}
\usepackage{graphicx}
\usepackage{wrapfig}
\usepackage{textcomp}
\usepackage{xcolor}
\usepackage{tabularx}
\usepackage{authblk}
\usepackage{svg}
\usepackage{listings}
\usepackage{enumitem}
\usepackage[font=footnotesize]{caption}
\usepackage{subcaption}
\usepackage{footnote}
\usepackage{array}
\newcolumntype{L}[1]{>{\raggedright\arraybackslash}m{#1}}
\newcolumntype{C}[1]{>{\centering\arraybackslash}m{#1}}
\newcolumntype{R}[1]{>{\raggedleft\arraybackslash}m{#1}}
\usepackage{varwidth}
\usepackage{amssymb}

\lstdefinestyle{lstStyleBase}{
	,basicstyle=\scriptsize\ttfamily 
	,floatplacement=tbp    
	,aboveskip=\smallskipamount 
	,belowskip=\smallskipamount 
	,lineskip=0pt          
	,boxpos=c              
	,showlines=false       
	,extendedchars=true   
	,upquote=true         
	,tabsize=2,           
	,showtabs=false       
	,showspaces=false     
	,showstringspaces=false 
	,numbers=none         
	,stepnumber=2         
	,numberfirstline=false 
	,numberstyle=\tiny\color{blue}    
	,numbersep=5pt        
	,numberblanklines=true %
	,numberbychapter=true %
	,captionpos=b         
	,abovecaptionskip=\smallskipamount 
	,belowcaptionskip=\smallskipamount 
	,linewidth=\linewidth 
	,xleftmargin=0pt      
	,xrightmargin=0pt     %
	,resetmargins=false   
	,breaklines=true      
	,breakatwhitespace=false 
	,breakindent=0pt     
	,breakautoindent=true 
	,columns=flexible     %
	,keepspaces=true      %
	,frame=lines         
	,framesep=3pt 
	,rulesep=2pt          
	,framerule=0.4pt      
	,language=C
}

\usepackage[
style=ieee,
backend=biber,
maxcitenames=1,
mincitenames=1,
uniquelist=false,
maxbibnames=999,
maxnames=1,
minnames=1,
doi=false,
url=true,
isbn=false,
]{biblatex}

\usepackage[linesnumbered,ruled,vlined]{algorithm2e}

\SetCommentSty{mycommfont}
\SetKwInput{KwInput}{Input}                
\SetKwInput{KwOutput}{Output}              

\makeatletter
\newcommand{\removelatexerror}{\let\@latex@error\@gobble}
\makeatother

\usepackage[%
textsize=footnotesize
]{todonotes}
\def\BibTeX{{\rm B\kern-.05em{\sc i\kern-.025em b}\kern-.08em
    T\kern-.1667em\lower.7ex\hbox{E}\kern-.125emX}}

\begin{document}

\font\myfont=cmr12 at 30pt
\title{\myfont Acceleration-as-a-$\mu$Service:\\ 
\Large{A Cloud-native Monte-Carlo Option Pricing Engine on CPUs, GPUs and Disaggregated FPGAs}}

\author[ ]{Dionysios Diamantopoulos, Raphael Polig, Burkhard Ringlein, Mitra Purandare, \\ Beat Weiss, Christoph Hagleitner, Mark Lantz, François Abel \\ IBM Research Europe, S\"aumerstrasse 4, 8803, R\"uschlikon, Switzerland \\ \textit {\{did, pol, ngl, mpu, wei, hle, mla, fab\}@zurich.ibm.com}}


\renewcommand{\baselinestretch}{1.1}
\maketitle
\renewcommand{\baselinestretch}{1.0}

\begin{abstract}

The evolution of cloud applications into loosely-coupled microservices opens new opportunities for hardware accelerators to improve workload performance. Existing accelerator techniques for cloud sacrifice the consolidation benefits of microservices. This paper presents CloudiFi, a framework to deploy and compare   accelerators as a cloud service. We evaluate our framework in the context of a financial workload and present early results indicating up to 485$\times$ gains in microservice response time.


\end{abstract}


\renewcommand{\baselinestretch}{0.9}

\section{Introduction}
Traditional Fortune 500 businesses prefer cloud services after realizing the cost benefits derivable from leasing cloud resources. These profits are empowered by the cloud computing model which entails two unprecedented advantages: \textbf{ scale} and \textbf{consumability}. 

The end of Dennard scaling and the slowdown of Moores' law have led to the use of increasingly specialized accelerators such as GPUs, TPUs, and FPGAs to improve cloud workload performance~\cite{10.1145/3317550.3321423}. In addition to using vendor accelerator offerings, major cloud providers are developing entirely new accelerators. For example projects like Microsoft's FPGA-based Brainwave or Google's TPU demonstrate the presence of economic incentives for cloud providers to invest in accelerated innovation, from the chip level to datacenter pods.

The consumability advantage refers to a broad class of technology layers that introduce new hardware and software innovations transparently to the users. Virtualization technology is the foundational technology layer for a cloud provider to offer processing, storage, and networking components without requiring coordination with their clients. It also enables transparent sharing of time and hardware allocations among clients with volatile needs. Virtualization offers a transparent vertically integrated solution to enable business models such as Infrastructure-as-a-Service (IaaS) and Platform-as-a-Service (PaaS) on public, private and on-premises clouds.

However, the new hybrid cloud, in which on-premises IT is combined with public and private clouds, necessitates architectural flexibility and cloud-native technologies to modernize applications for improved return-on-investment and faster time-to-market. 

Among the cloud native architectural proposals~\cite{etzkorn2017introduction}, the microservices architecture is one of the accelerating trends these days~\cite{10.1145/3183628.3183631}. The concept is depicted in Fig.~\ref{fig1} with the transformation of classical layered integration based on virtualization~\cite{7217792}, legacy containerization technology~\cite{7922500} and  middleware~\cite{etzkorn2017introduction}, into a cloud-native middleware stack.  This switching is a key enabler for end-users to consume the stack at large scale and independently of the type of cloud environment, i.e. public, private, or on-prem [5].

Financial service companies are also considering microservices as a vital asset for their digital transformation~\cite{mservices_market_report}. Furthermore, the combined use of accelerators and microservices by fintech professionals is expected to deliver significant speedup advantages.

\textbf{The contributions of this paper are: (1)} A cloud-native framework, named \texttt{CloudiFi}, that enables microservices to access hardware accelerators. \textbf{(2)} A best-effort work-in-progress evaluation of this framework with a representative quantitative finance workload.  Our results indicates that a network-attached mid-range FPGA accelerator yields better end-to-end (E2E) response time and scalability, for two load ranges, compared to latest generation CPUs and GPUs.

{\setlength{\belowcaptionskip}{-2ex}
\begin{figure}[t]
\includegraphics[width=0.77\textwidth]{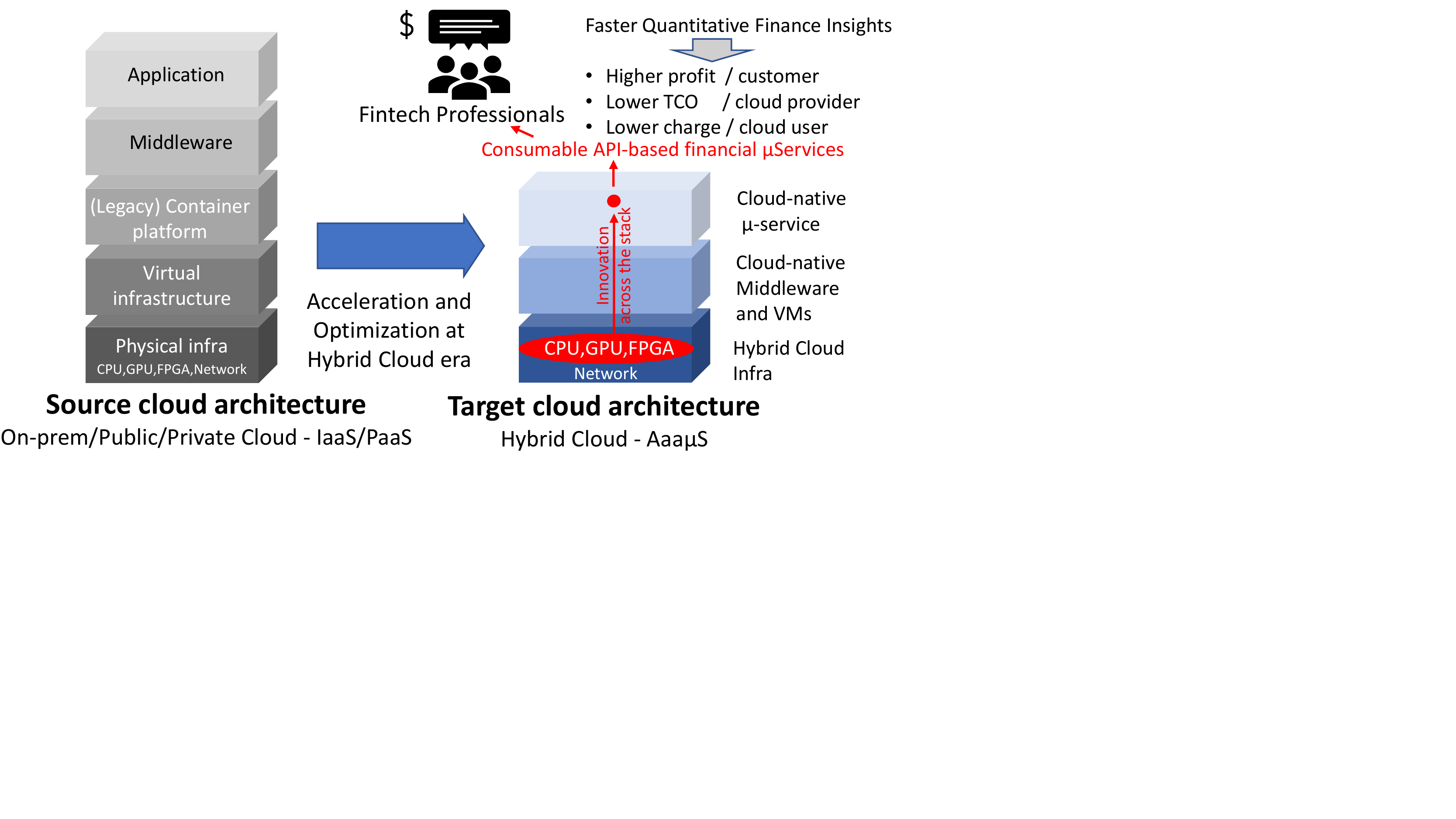}
\vspace*{-113pt}
\caption{Motivation of this work-in-progress: Accelerators in the hybrid cloud.}
\vspace{-7pt}
\label{fig1}
\end{figure}
}

\section{Framework for accelerating microservices}

{\setlength{\belowcaptionskip}{-4ex}
\begin{figure*}[t]
\includegraphics[width=1\textwidth]{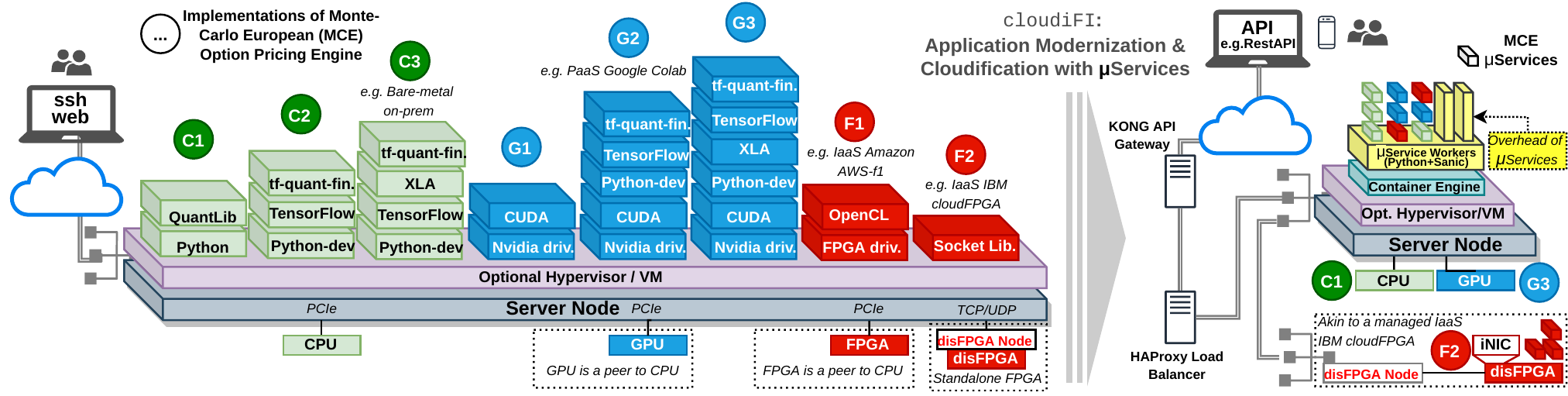}
\caption{Overview of \texttt{CloudiFi}: Turning an Infrastructure/Platform into a cloud $\mu$Service.}
\label{fig5}
\end{figure*}
}

\textbf{A. Quantitative Finance Case Study for Cloud:} To demonstrate the feasibility of \texttt{CloudiFi} we require a candidate application that benefits from Acceleration-as-a-$\mu$Service (Aaa$\mu$S). For this we selected an application employed in quantitative finance, referred to as 'Monte-Carlo European Option Pricing' (MCE). The analysis of MCE is underpinned by a Monte-Carlo simulation that is computationally intensive. Typically, this costly analysis is periodically performed on production clusters.
MCE is used for option pricing where numerous random paths for the price of an underlying asset are generated, each having an associated payoff. These payoffs are then discounted back to the present and averaged to get the option price. It is similarly used for pricing fixed income securities and interest rate derivatives. But the Monte Carlo simulation is used most extensively in portfolio management and personal financial planning. In this case, the pricing of options needs to be obtained in real-time, e.g. from exchanges or other fintech parties, in order to inform the next decision (e.g. sell/buy stocks etc.). The speedup offered by hardware accelerators can be leveraged to develop a faster application fit for use in real-time settings. The \texttt{CloudiFi} framework, which is depicted in Fig.~\ref{fig5} and discussed in subsection~\textbf{II.C}, suits such applications because minimal changes to the production cluster are needed and the acceleration required for the analysis is obtained as a service from a remote host.

\begin{wrapfigure}{r}{0.15\textwidth}
  \begin{center}
    \vspace{-10pt}
    \includegraphics[width=\linewidth]{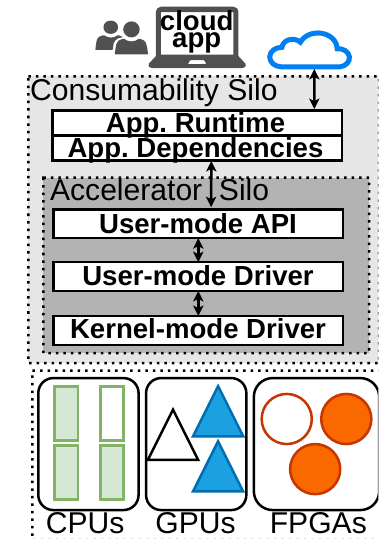}
  \end{center}
  \vspace*{-1em}
  \caption{A cloud application, tightly coupled silos, and HW.}
  \label{fig6}
  \vspace{-1em}
\end{wrapfigure}

\noindent \textbf{B. Cloud-native considerations:} 
A HW-accelerated financial workload on cloud is typically associated with a software stack of application dependencies, a runtime, proprietary protocols and kernel bypass techniques. Such a "cloudification stack" is depicted in Fig.~\ref{fig6} and constitutes what we refer to as a silo. First, the \textit{accelerator silo}, as already introduced by~\cite{10.1145/3317550.3321423}, is formed of tightly coupled vertical layers, that communicate either through proprietary interfaces, and/or using low-level mechanisms and essentially provide only one user-mode API interface. This is typically consumed by cloud users within a VM on IaaS/PaaS setups on public, private and on-premises clouds.

Next, we define a \textit{consumability silo} as a silo that includes the previous silo together with all the library dependencies that allow a standalone application to be executed on a cloud instance. E.g., if the cloud instance is a VM of a public IaaS, and our application is written in Python, then the consumability silo refers to the software stack required in this VM, from the base OS to the Python dependencies/runtime of this application, alongside the accelerator-specific device drivers.

%

In this paper, we introduce a novel framework to break the consumability silo between acceleration infrastructure on CPUs, GPUs and FPGAs and software microservices on cloud servers. We defined a cloud architecture where microservices can access various accelerators for function offloading and acceleration. 

\noindent \textbf{C. The \texttt{CloudiFi} Framework:} We developed the \texttt{CloudiFi} framework following the \textit{Twelve-Factor App} microservices best practices, methodologies and guidelines~\cite{12factor}. The source architecture is composed of a compute node connected to different accelerators, as depicted in the left part of Fig.~\ref{fig5}. For every accelerator we select different implementations for our financial workload, as discussed later in Section~\ref{sec:mce}. Every implementation is associated with an \textit{consumability silo}. The right part of Fig.~\ref{fig5} shows the target architecture of the \texttt{CloudiFi} framework. We briefly detail the main components of the target architecture, as depicted in Fig.~\ref{fig7}.
The option pricing requests are issued from the cloud users using a \textbf{RESTful API}, an API exposed via HTTP endpoints. The \textbf{KONG API Gateway} is a service that enables load balancing and scales the overall service throughput.

{\setlength{\belowcaptionskip}{-1.8ex}
\begin{figure}[h]
\includegraphics[width=0.5\textwidth]{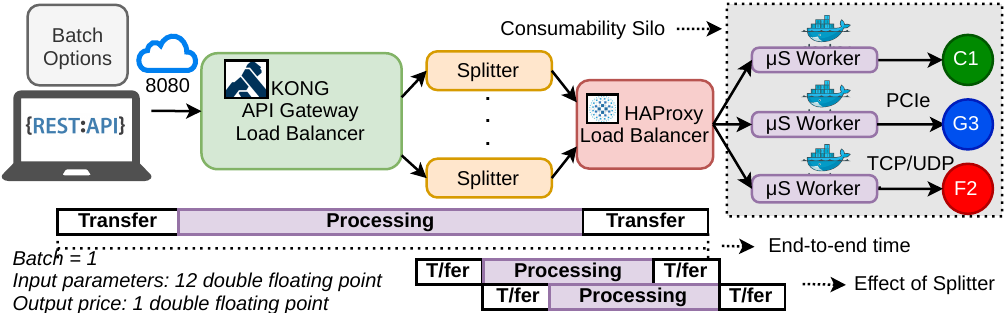}
\vspace{-10pt}
\caption{The architecture of \texttt{CloudiFi} framework.}
\label{fig7}
\end{figure}
}
The \textbf{Splitter} is a component that uses the same REST API with the clients but splits the request into smaller requests in order to parallelize the processing time, as depicted in the lower part of Fig.~\ref{fig7}. The \textbf{HAProxy} is the central load balancer that distributes the smaller batches to the workers.
Finally, the \textbf{$\mu$Service Workers} are the processing threads of a hosting node that serve the cloud requests. The workers are based on Python runtime with the Sanic web framework on top. This allows the usage of asynchronous, concurrent and non-blocking tasks. We use the \textit{blueprint} object of Sanic that routes the incoming request to the code of every $\mu$Service. We implemented a python wrapper for every accelerator as a \textit{blueprint} object. The workers are executed in an isolated environment using Docker containers, enabling increased stability, system robustness and cloud-native development processes~\cite{12factor}. 

\section{MCE characterization}
\label{sec:mce}
Before assessing the performance of a set of accelerators in \texttt{CloudiFi}, we first benchmark them as standalone applications. Our compute node is composed of a server node connected to different accelerators, as depicted in Fig.~\ref{fig5} (left part). The server is used for processing the MCE workload. We selected the following implementations: \textbf{C1)} a CPU implementation using the QuantLib library~\cite{quantlib} (Python), \textbf{C2)} a CPU implementation using the TensorFlow (TF) Quant Finance (tfquant) library from Google~\cite{tfquant}, \textbf{C3)} the same as C2, but using the XLA compiler~\cite{xla}, \textbf{G1)} an in-house GPU implementation of MCE using the CUDA library, \textbf{G2)} the GPU version of C2, \textbf{G3)} the GPU version of C3, \textbf{F1)} an FPGA implementation on AWS using the Vitis Quantitative Finance library~\cite{vitis}, \textbf{F2)} F1 using a disaggregated FPGA node (disFPGA)\cite{8071053}.

All of these best-effort implementations are state-of-art (SoA) MCE workloads, to the best of our knowledge. To explore which are suitable for cloudification, we first benchmark them as standalone applications, avoiding any overheads related to microservices. All CPU and GPU implementations are executed on the same on-prem server node. The F1 is evaluated on an \textit{AWS f1.x2large} instance (Xilinx VU9P FPGA tightly attached to Intel Xeon E5-2686 v4 Broadwell CPU). The difference between F1 and F2 is that the former is a typical FPGA attachment to the CPU over PCIe, similar to how GPUs are attached to CPUs, i.e. as peer co-processors. In the latter, the FPGA is decoupled from the CPU of the server and directly connected to the network. This is enabled by implementing an integrated Network Interface Card (iNIC) into the FPGA logic, as described in~\cite{8071053}. Our benchmarking results are shown in Fig.~\ref{fig2}. For every measurement we report \textit{cold} and \textit{hot} times. Cold accounts for the first time of execution. Hot refers to the geometric mean of 100 runs after the first time of execution. 

Due to CPU/GPU cache misses, we observed differences on the order of 26\%/42\% for the C1/G1 cold versus hot implementations respectively. However in the other CPU and GPU implementations, that employ the \textit{tfquant} library, we observed a difference of one to two orders of magnitude, which is attributed to the graph optimization time of TF. For F1 we also measured a difference of 20.1$\times$ due to the overhead in opening the FPGA device, for the first time, through the OpenCL-based Xilinx  runtime.
The F2 implementation exhibits a difference of 24\% 
that is related to whether the TCP/UDP connection is initialized (cold) or the TCP/UDP socket is already open (hot). 
In general, we observed performance deviations of up to 1380$\times$, depending on the accelerator,  compiler and HW/SW stack readiness (hot-cold time). Here, due to the low-latency requirement of our financial workload, we examine stateless microservices, i.e. processes that do not save or reference information about previous operations. In contrast, stateful microservices offload persistence to the host or use highly available cloud data stores to provide a persistence layer. Consequently, for \texttt{CloudiFi}, we selected the implementation with the lowest cold execution time, for every accelerator, i.e. C1-cold, G3-cold, F2-cold for CPU, GPU, FPGA, as depicted in Fig. \ref{fig2}.

{\setlength{\belowcaptionskip}{-2ex}
\begin{figure}[t]
\includegraphics[width=0.7\textwidth]{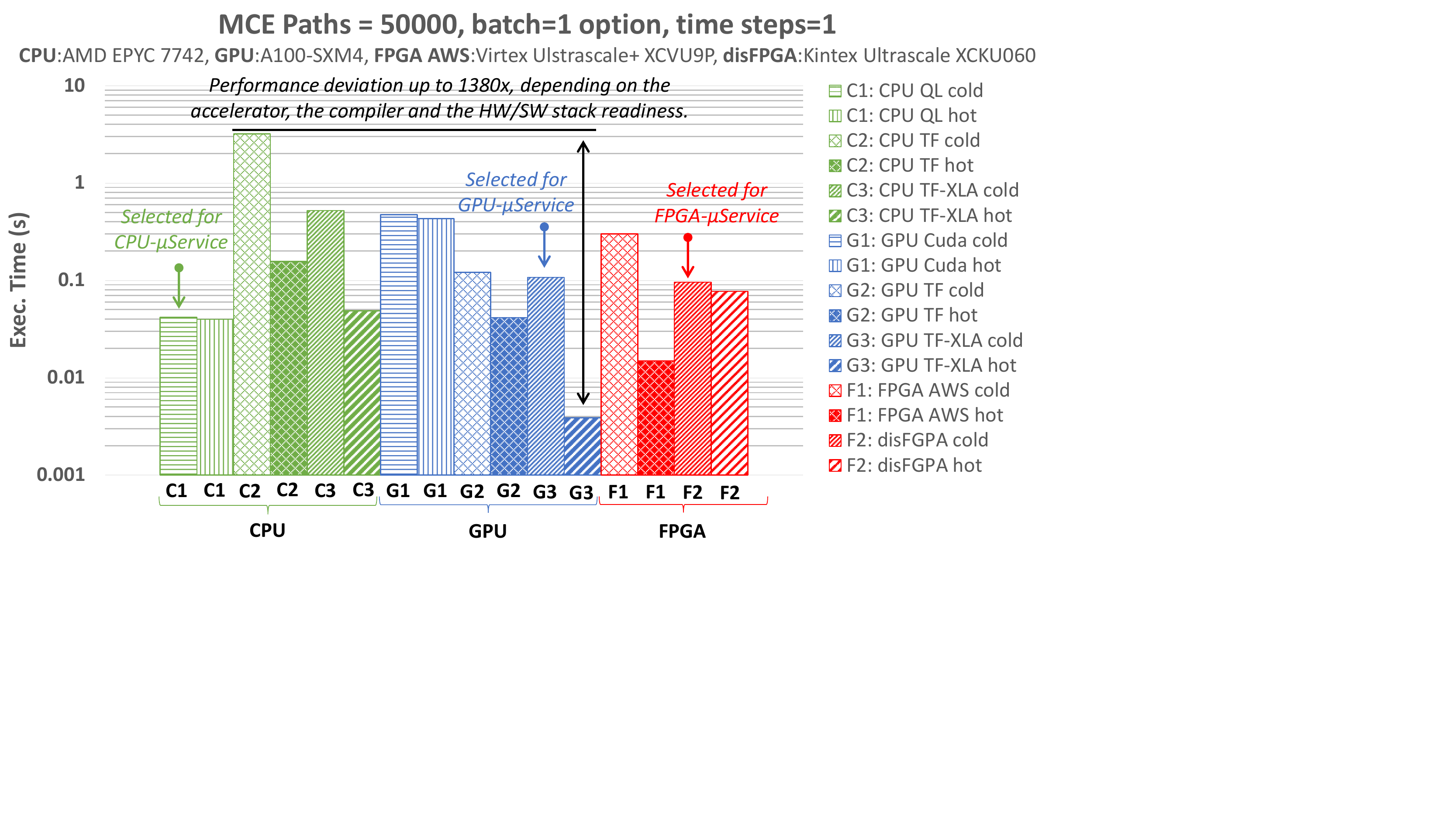}
\vspace{-80pt}
\caption{MCE characterization by benchmarking SoA finance libraries.}
\label{fig2}
\vspace*{-0.5em}
\end{figure}
}

\section{Preliminary Experimental Results}
We implemented \texttt{CloudiFi} on a dual-socket AMD EPYC 7742 CPU (7nm, 64-Core/128-threads, 2.25GHz, TDP 225W), 1TB RAM, with a SXM4-based A100 GPU (7nm, 1095MHz, 40GB HBM2e 1,555GB/s, TDP 400W) connected via PCIe 4.0x16 (32GB/s), running Ubuntu 20.04 LTS. An identical CPU-only system is configured to send batch requests of options over the network, using RestAPI \textit{GET} commands. The systems are connected to the network with 200Gb/s HDR ConnectX-6 InfiniBand adapter cards. The network is also accessed from a remote disFPGA node, 
from which we configured one network-attached FPGA device (Kintex Ultrascale XCKU060, 20nm, 156MHz) with seven parallel MCE cores (\cite{vitis}) and one 10Gb/s UDP processing engine. These parallel cores can process a batch of seven options in parallel, while resulting in FPGA utilization of 97\% LUTs, 82\% FFs, 65\% DSPs and 56\% BRAMs. We run a concurrent peer-to-peer client-server application. We analyze two different scalability modes and compare i) the processing of the worker (purple bar in the lower part of Fig.~\ref{fig7}) and ii) the end-to-end $\mu$Service response time (purple and white bars of Fig.~\ref{fig7}), using different accelerators.

We first compare the $\mu$Services scalability with respect to the number of Monte-Carlo simulation parallel paths. As depicted in Fig.~\ref{fig3}(a), 
the \textit{disFPGA} response time (both Proc. and E2E) remains almost stable across the entire load range. In contrast, while the GPU processing time is the lowest of all implementations, it exhibits an E2E time up to three orders of magnitude higher, as a result of the high \textit{cold-time}. The existing accelerator consumability options forces cloud vendors to dedicate physical hardware to VMs~\cite{10.1145/3317550.3321423}. As a result, they sacrifice the consolidation benefits of microservices that are fundamental to the emerging cloud-native workloads. The baseline CPU is fast for few paths but scales quickly with  E2E time. Fig.~\ref{fig3}(b), compares the processing time and E2E time for different implementations, under the fixed selection of \textit{500,000} paths, for a load range of [1, 10, 100] batch size. 
We observed a similar trend again, which is highlighted by the trend lines. In addition, as depicted in Fig.~\ref{fig3}(c), the container image size of the \textit{disFPGA} $\mu$Service workers is significantly lower than the CPU and GPU versions, due to the smaller number of necessary libraries (Fig.~\ref{fig5}).

{\setlength{\belowcaptionskip}{-2ex}
\begin{figure}[t]
\includegraphics[width=1.3\textwidth]{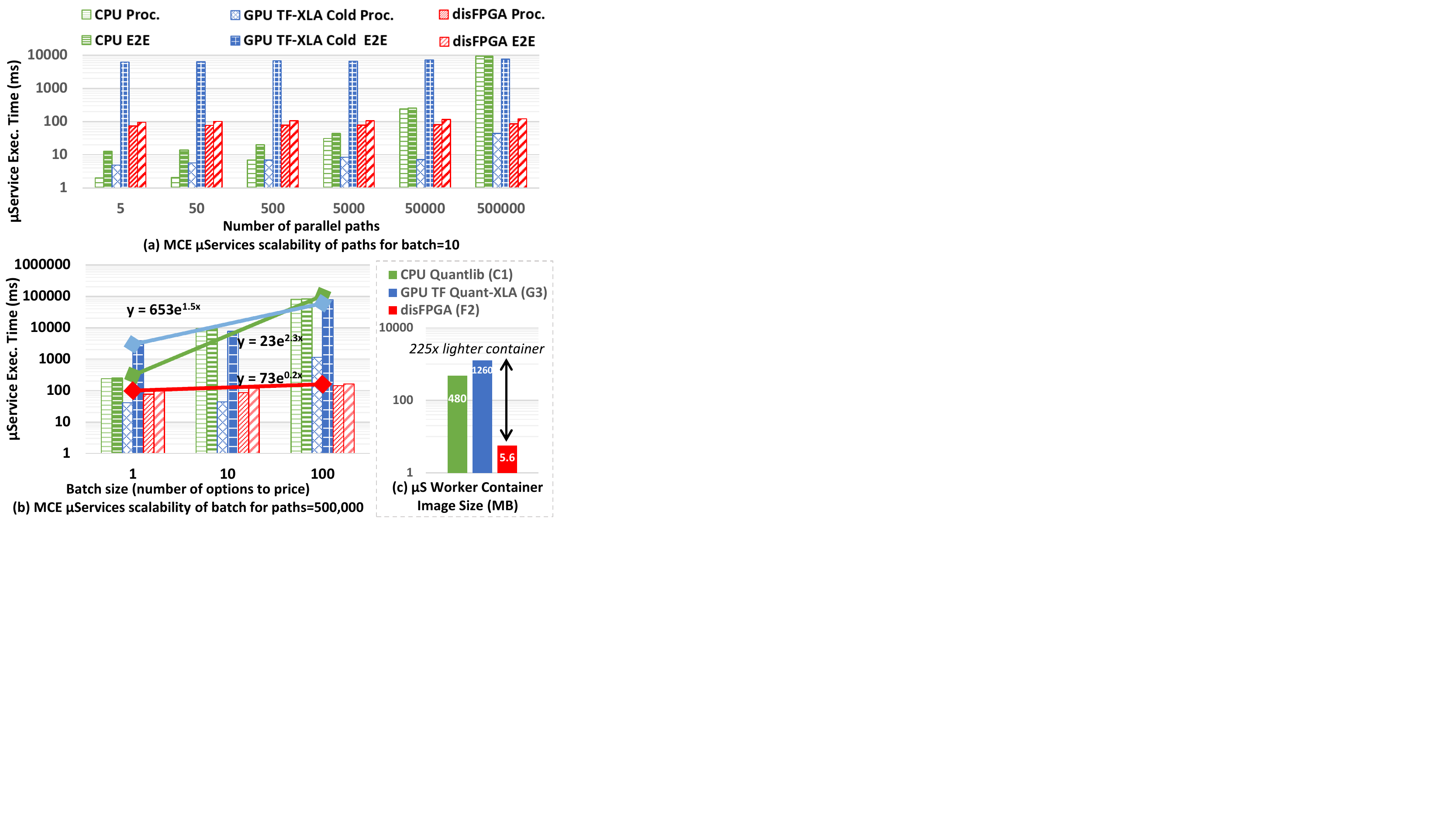}
\vspace*{-150pt}
\caption{\texttt{CloudiFi}: Preliminary experimental results of a MCE Aaa$\mu$S.}
\label{fig3}
\end{figure}
}

\section{Conclusion}
\texttt{CloudiFi}, our work-in-progress on hardware-accelerated cloud-native microservices, significantly decreases 
the E2E $\mu$Service response time, by up to 485$\times$, using network-attached disaggregated FPGAs. At the same time it highlights the consideration of statefull $\mu$Services for PCIe-attached GPUs, since in our setup, the excessive overhead of wrapping a GPU over a cloud-native service - i.e. the consumability silo - supplants the unprecedented acceleration capability of even the latest generation GPUs. Our framework is orthogonal to the underlying hardware accelerators or workload domains and in the future we are looking forward to implement other state-of-art accelerators for microservices and state-of-art benchmarks~\cite{10.1145/3297858.3304013}.

\renewcommand{\baselinestretch}{0.85}
\printbibliography

\end{document}